# The astronomical orientation of the urban plan of Alexandria

*Luisa Ferro and Giulio Magli*
*Faculty of Civil Architecture, Politecnico di Milano, Italy*

*Alexander the Great founded Alexandria in 331 BC. Examining the topography of the city today allows the identification of the essential elements of the original urban system, and shows that the choice of the site was mainly due to religious and symbolic reasons. As a matter of fact Alexandria was the prototype of a series of Hellenistic towns designed as "king's towns" aiming to make the divine power of their founder explicit. This being the motivation, we examine the orientation of the orthogonal grid, which was based on a main longitudinal axis, and show that this axis is orientated to the rising sun on the day of birth of Alexander the Great. At the time of foundation, "king's star" Regulus was rising, as well, along the same direction. A series of arguments, based on the topography of the town and on the comparison with later Hellenistic towns and monuments - such as Seleucia on the Tigris and the Nemrud Dag - is given to support the thesis that this orientation was not due to chance but was, instead, a deliberate one.*

# 1. Introduction

During the 4th century several debates concentrate on the idea of "ideal town". Socrates, in the *Gorgia,* anticipates a new conception of town which goes beyond mere practical principles, and Plato repeatedly prefigures the birth of the ideal city (e.g. *Laws* IV, V). The inspiring principles are based on harmony as related to the laws and the divine, reflected in the mathematical rigour of the design of the "Hippodamean" city plan (Castagnoli 1971, Shipley 2005). Slightly later, Aristotle will combine such ideas with a sort of "biological" way of thinking linked to the concept of variety and several necessities, stating for instance the need for springs and fortified walls.

Alexander the Great founded Alexandria in 331 BC (Bagnall 1979). The foundation of Alexandria can be set as the apex of such debates, as well as the beginning of a series of new towns, the ones of the Seleucids (Wheeler 1968). The city becomes an explicit representation of the power of its divine founder, the rigorous order of its plan being a reflection of the "cosmic" order, in compliance with the "orthogonal grid" principles. The orthogonal grid of Alexandria can still be perceived, and forms the basis for an ongoing project of reassessment of antiquities into a coherent architectural scheme of fruition (Ferro and Pallini 2008, Torricelli 2010a, 2010b). The town was wholly designed from the very beginning in all its details, with a relatively huge perimeter, and was divided into five areas named after the first five letters of the Greek alphabet. The original matrix *route* was conceived on the basis of a longitudinal axis, later called *Canopic Road;* the most important transverse axis was a dyke (*Eptastadion*) connecting the mainland with the isle of Pharos. The Canopic Road played the role of an "extended centre", a wide, longitudinal open space, with the main buildings distributed along it, thus avoiding the idea of a "central point" as the focus of the urban plan. The first to put in evidence such a "longitudinal" character in the original project of Alexandria was 19[th] century astronomer Mahmud Bey Al-Falaki (1861). Later excavations along the modern street showed that the Canopic Road was actually deeply etched in the rock subsoil (Breccia 1914). The axis is thus a peculiar characteristic, a sort of icon in the foundation of the city, and as such it is an independent architectural unity (Mumford 1967, Caruso 1993, Ferro 2010). It will be repeated in later town projects and, in particular, in the design of Seleucia on the Tigris, as we shall see in further details later on.

In spite of what is emphatically reported by Plutarch in his *Life of Alexander* (26, 2-3) and by Diodorus Siculus (17, 52) the site where the newly founded town was built did *not* have special characteristics of suitability. In particular, the city was planned in a strip enclosed between the sea to the north and west, the marshy lands of the Canopus mouth of the Nile to the east, and Mareotis Lake to the south, in contrast with many of the healthy criteria of Alexander's tutor, Aristotle (Bernand 1995). Further, a series of preliminary works was required in order to construct in the desired place: Alexander's *Romance* reports about the existence of 12 channels which had to be dried in order to cover them with streets, and excavations have in fact shown the existence of at least 3 of such channels. Another characteristic (to be discussed later on) which clearly conflicts with utilitarian principles is that the orthogonal grid seems to be *not* conformal to the characteristics of the landscape.

All in all, we are led to consider the foundation of Alexandria as a truly symbolic act, inspired by "religious" criteria and aiming at the celebration of Alexander's power and divine nature (Ross Taylor 1927). Such a foundation probably was in compliance with the – already old – ritual: indeed the founders visited an oracle before starting their enterprise. Typically, it was the oracle at Delphi. Alexander, however, visited the most important oracle in Egypt, the Ammon oracle at Siwa, and this probably occurred just before the foundation (Bradford Welles 1962). The founder played the role of ancestor for the town as a whole, and was to be buried "in the center of the city" (Detienne 1998) though the burial of Alexander – as is well known - has not yet been found.

Among the possible symbolic aspects related with foundation and to be considered in the analysis of a town's project is, of course, orientation. In particular, it has been repeatedly suggested that several *Roman* towns, whose orientation does not conform to peculiar features of the landscape, were oriented in accordance with astronomical, rather than utilitarian, criteria. This fact has been recently investigated in a systematic way in the case of Roman towns in Italy (Magli 2008). The present paper aims at investigating the same issue in the case of the orientation of Alexandria.

**2. Orientation of the urban plan of Alexandria**

As mentioned above, the rectangular grid of Alexandria was based on the so called *Canopic road* which crossed the city leading to the Canopic mouth of the Nile and the Canopus (today Abukir) bay. At the opposite ends of the street two main gates were located; at least since the work of Achilles Tatius (early second century AD) the east and west gates were called Gate of the Sun and Gate of the Moon respectively (Haas 1997). The Canopic road bears an azimuth of 65°15' ± 30' [1]. The horizon to the east extends towards the Abukir bay and was therefore flat in ancient times; the same holds to the west. Possibly, the only favourable point for the surveyors of the newly founded town was the "hill" of the Serapeum, located to the south-west of the town, but even in this case the elevation is negligible (some 15 meters) so that the horizon on the sea can be considered as flat on both sides.

In 331 BC the azimuth of the rising sun at the summer solstice - occurring on June 28 (all the dates in this paper are Julian) - was 62° 20' (today it is slightly displaced due to the variation in the obliquity of the Ecliptic). It can of course be said, therefore, that the orientation of Alexandria axis at 65° 15' is "solar" in that the sun was (and is) rising along this direction twice a year. The dates are July 24 and the symmetric date, in relation to the summer solstice, June 2. The range - one degree wide - centred on azimuth 65° 15', was spanned by the rising sun in a period of a few days before and after this date, respectively. It is the aim of the present paper to defend the idea that this orientation was deliberate.

First of all, one could hypothesize a very rough solstitial alignment. However, the difference of some 3 degrees – and therefore an error of 3 degrees in determining the direction of the rising sun - looks quite exaggerate, both for the Egyptian and for the Greek standards of the period (Magli 2009). We propose here a quite different possibility, namely that the city was orientated to the rising sun on the day of the birth of Alexander the Great. Alexander was indeed born on July 20, 356 BC, and in the 4th century BC the sun was rising at Alexandria on that day at an azimuth 64° 30', only 45' less than our best estimate for the azimuth of the Canopic Road. The Julian date of birth of Alexander, however, has nothing to do with the calendar in use in that earlier period, and therefore our proposal requires a careful discussion.

It is most probable that the calendar used by the planners of Alexandria was not the Egyptian solar ("civil") calendar running 365 days per year. This calendar lost around 6 days in relation to the sun cycle between 356 and 331 BC. It was, rather, the luni-solar Greek calendar to be in use (Hannah 2005). According to ancient sources such as Plutarch's, Alexander was born on the sixth day of *Hecatombaeon*, the first month of the year. New year's day was the day of the first new moon after the summer solstice, and, in 356 BC, this occurred on July 14, leading to July 20 for Alexander's birth. [2] Due to the length of the synodic month (about 29.53 days) however, the date of the new moon after the summer solstice varies from year to year, so that *Hecatombaeon* 6 wanders through different solar dates as well [3]. In other words, the date of birth of Alexander is not a fixed day of any solar calendar. It is, however, also true that Greek astronomers were perfectly able to trace the date of the new moon back in time. Indeed the Metonic cycle (stating that 19 tropical years are needed to complete 235 synodic months) was known to them since the 5th century. The day of Alexander's birth was, together with the foundation of the city (Tybi 25, which fell on April 7 in 33I BC) the most important festivity of the town, celebrating Alexander as a living God. As any festivity fixed according to the moon (e.g think of Christian Easter) the date varied from year to year, as did many other pre-existing festivities in the Greek world. Due to the waving of the lunar calendars and to differences between different local calendars, the Greeks elaborated astronomical methods to act as harbingers for relevant festivities (Hannah 2005, 2009; Salt and Boutsikas 2005). In this respect, the alignment to the rising sun in the companion solar date might have been used as a solstitial marker – and therefore as a "correct new moon indicator" - occurring a number of days (26) close to a lunar month *before* the summer solstice.

In addition, heliacal rising of stars was used as harbinger of important festivals, and the Alexandria alignment was operational also in this sense. Surprisingly indeed, the star associated with kingship since Babylonian times, that is "King's Star" Regulus (alpha-Leonis), was at that time rising at the very same azimuth (65° 20' at altitude of one degree, appropriate for the visibility of a first magnitude star) and had heliacal rising very near to July 20 (the precise date of Heliacal rising of a star depends on many factors and cannot be defined with high precision - see Schaefer 1986).

**3. Seleucia on the Tigris and the Nemrud Dag**

When a deliberate astronomical alignment is proposed, it is of course fundamental to investigate on the possibility of a mere coincidence. If the sample under exam is wide enough, a statistical analysis can be applied to evaluate the probability of casual alignments (Ruggles 2005). A similar analysis has been recently carried out by one of us in the case of Roman towns in Italy (Magli 2008). Unfortunately, Alexandria is unique in relation to the previously founded Greek towns since, as mentioned above, it is not the result of a colonization process, but it is, rather, the first example of a "city of the king". To understand if the alignment is likely to be intentional we can, however, at least analyse if the same alignment occurs in other cases within the same cultural context. To this aim, we will discuss two particularly interesting examples. The first, the orientation of Seleucia on the Tigris, is original to the present paper, while the second is due to Belmonte and García (2010).

As recalled above, Alexandria has been considered as a sort of model town during Hellenism, and some cities are known to have been inspired by its plan. The most important of such towns is perhaps Seleucia on the Tigris, which was founded in 300 BC by Seleucus I Nikator (305-281 BC) (Hadley 1978). The site is not far from Babylon, where Alexander died on June 10, 323 BC and where the first residence of the Seleucids was established. The king decided to found a new capital in a site placed in a better location than Babylon in the contemporary network of roads. As much as for Alexandria, the foundation of the town is surrounded by several legends. Seleucus himself was deified as son of Apollo and therefore, once again, we see a process of identification of the newly founded town with the divine power of its founder. If Alexandria was planned to be "similar to a chlamys", as Diodorus says, the form of Seleucia resembled an eagle, at least according to Pliny. Be it as it may, the topography of Seleucia is manifestly inspired by that of Alexandria, with a main longitudinal road and a regular urban matrix nested on such an axis (Mumford, Caruso, Ferro 2010, Gullini 1972, Hopkins 1972, Messina 2007*)*. The city lies on a flat plain and therefore the planners had full liberty in orienting the town axis, according to the bank of the river. Nevertheless, the "Canopic" axis of Seleucia bears an azimuth of 295° ±30' (data extracted from archaeological maps and controlled on satellite images) which is therefore, within the errors, the symmetric azimuth with respect to the meridian of Alexandria's Canopic road. Would Alexandria and Seleucia be at the same latitude, the sun with a flat horizon would set in alignment with the longitudinal axis at Seleucia the very same days as at rising in Alexandria. Due to the slight difference in latitude the sun was actually setting along this direction on the days around July 27, with a slight displacement, but in any case still very close to the date of birth of Alexander. As well as in Alexandria, there is also a close concordance with Regulus, whose setting occurred approximately at azimuth 294° 40'. It is therefore tempting to speculate that Seleucia, being inspired by Alexandria both symbolically and practically, but being also close to the place of death of the revered king, was orientated towards the same astronomical targets, but at their settings.

A second interesting example is, in chronological order, the magnificent funerary monument of Antiochos I, King of Commagene, at Mount Nemrud (Belmonte and García 2010). The monument, constructed in the first half of the last century BC, includes the famous "lion horoscope", depicting Mars, Mercury, Jupiter and the crescent moon in Leo, the constellation of Regulus (Neugebauer and Van Hoessen 1959). The two terraces of the monument are orientated to the solstices, but the huge plinths holding the colossal statues in the eastern terrace point to sunrise on 23 July (and 22 May; Gregorian and Julian dates practically do not differ). According to Belmonte and Garcia, the 23$^{rd}$ of July coincides with the date of the celebration of Antiochos's ascent to the throne mentioned in the inscriptions of the monument. The coincidence with Alexandria is really striking, considering that Antiochos makes explicit reference to Alexander the Great as ancestor in the inscriptions on the monument. Moreover, precisely as in Alexandria, the terrace also points to the rising of Regulus, which also occurred around the 23rd July at the Nemrud latitude during the reign of Antiochos I.

## 4. Conclusions

The dissertations presented above definitively point, at least in our view, to the conclusion that Alexandria was deliberately orientated towards the rising sun on the day of birth of Alexander and the rising of the king's star Regulus. As a final observation, we may notice that the orientation of the Alexandria grid was not dictated by topographical reasons. Indeed, the natural choice would have been to trace the longitudinal axis parallel to the shoreline, which has a mean azimuth around 49° (the ancient shoreline of Alexandria was very similar to the present one, while extended submerged areas exist in the Abukir bay, see Goddio and Bernand 2005). Interestingly, when Roman architects sat about to construct the Caesareum (probably founded by Cleopatra in honour of Mark Antony and later dedicated to Caesar Augustus), they apparently felt the "failed" orthogonality of the grid to the main shoreline direction as a disturbing fact, because planning the temple in conformity to the existing grid did not give a satisfactory view of the monument from the port and therefore to the people arriving from the sea. So they decided to break the symmetry of the grid: the temple is now lost but its position has been reconstructed on the basis of the obelisks which were still standing in Alexandria in the 19th century, and – at least according to McKenzie (2008) - the front formed an angle of (roughly) 15° (i.e. 64-49) with the pre-existing longitudinal axis.

   To conclude, the examples of orientation to Regulus and to the sun at the end of July coming from the Nemrud Dag and from Seleucia are clear hints at the existence of a traditional pattern of orientation first established in Alexandria. Alexander the Great confirms himself, once again, as "marking a major devide in the broad history and archaeology ideas" as Sir Mortirmer Wheeler once said.

**Acknowledgements**

The present work arises within a vast research project (entitled *Archaeology and Architecture*) devoted to the enhancement of the Archaeological Areas of Alexandria, co-coordinated by Angelo Torricelli whose constant help and encouragement is gratefully acknowledged. The project includes a collaboration with the University of Torino, the Alexandria & Mediterranean Research Center, the Department of Architecture of Menofeya University, and the Italian Archaeological Mission at Alexandria coordinated by Paolo Gallo. The current project mission entitled "Kosa Pasha Fort, Abuqir" is operating under an International Protocol of scientific collaboration with the Supreme Council of Antiquities (SCA) of Egypt. Architect Viola Bertini and Elena Ciapparelli, who have worked on images, and students Marina Bianconi and Valentina Sala are also gratefully acknowledged.

CARTE DE L'ANTIQUE ALEXANDRIE ET DE SES FAUBOURGS
Dressée sur les ordres de S.A. le Vice-Roi d'Égypte à l'aide de fouilles nivellements et autres recherches par MAHMOUD-BEY, Astronome de Son Altesse
Fait en 1866

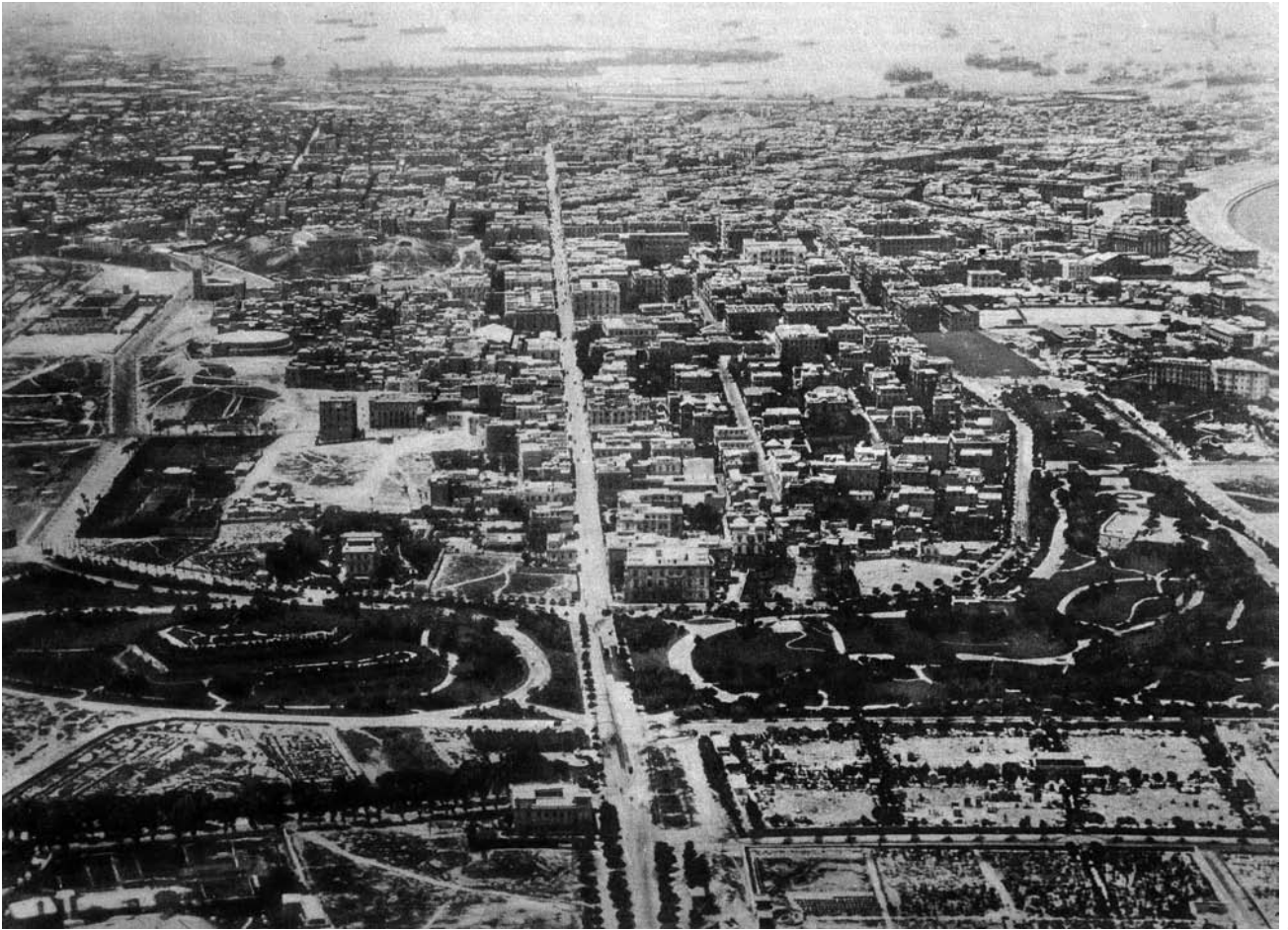

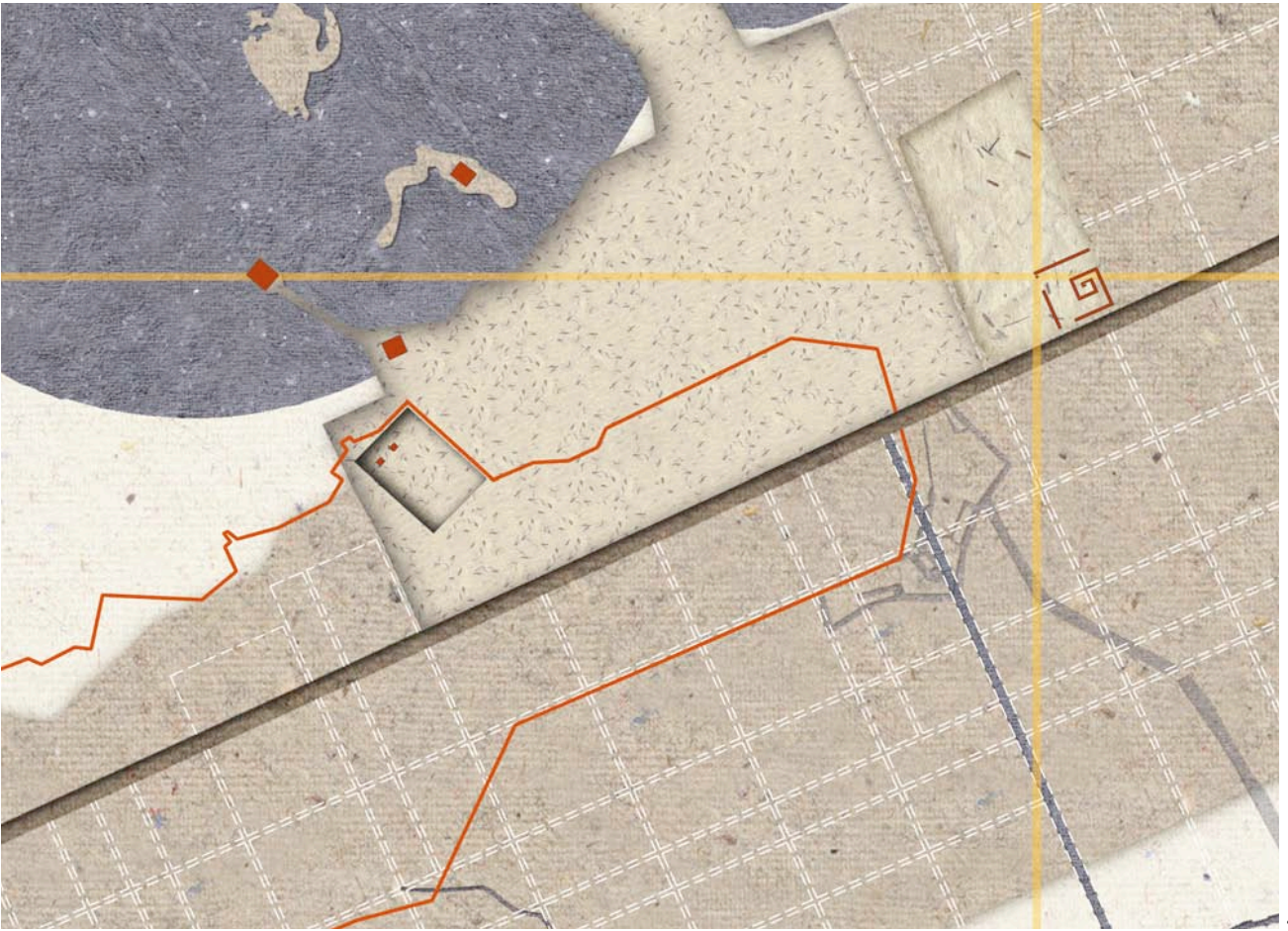

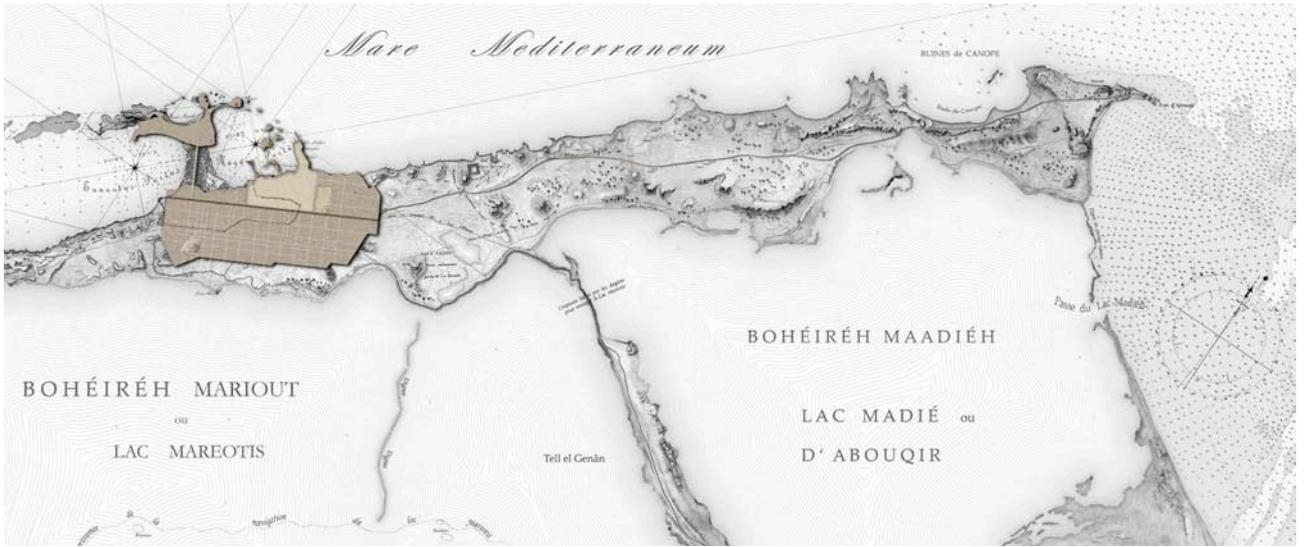

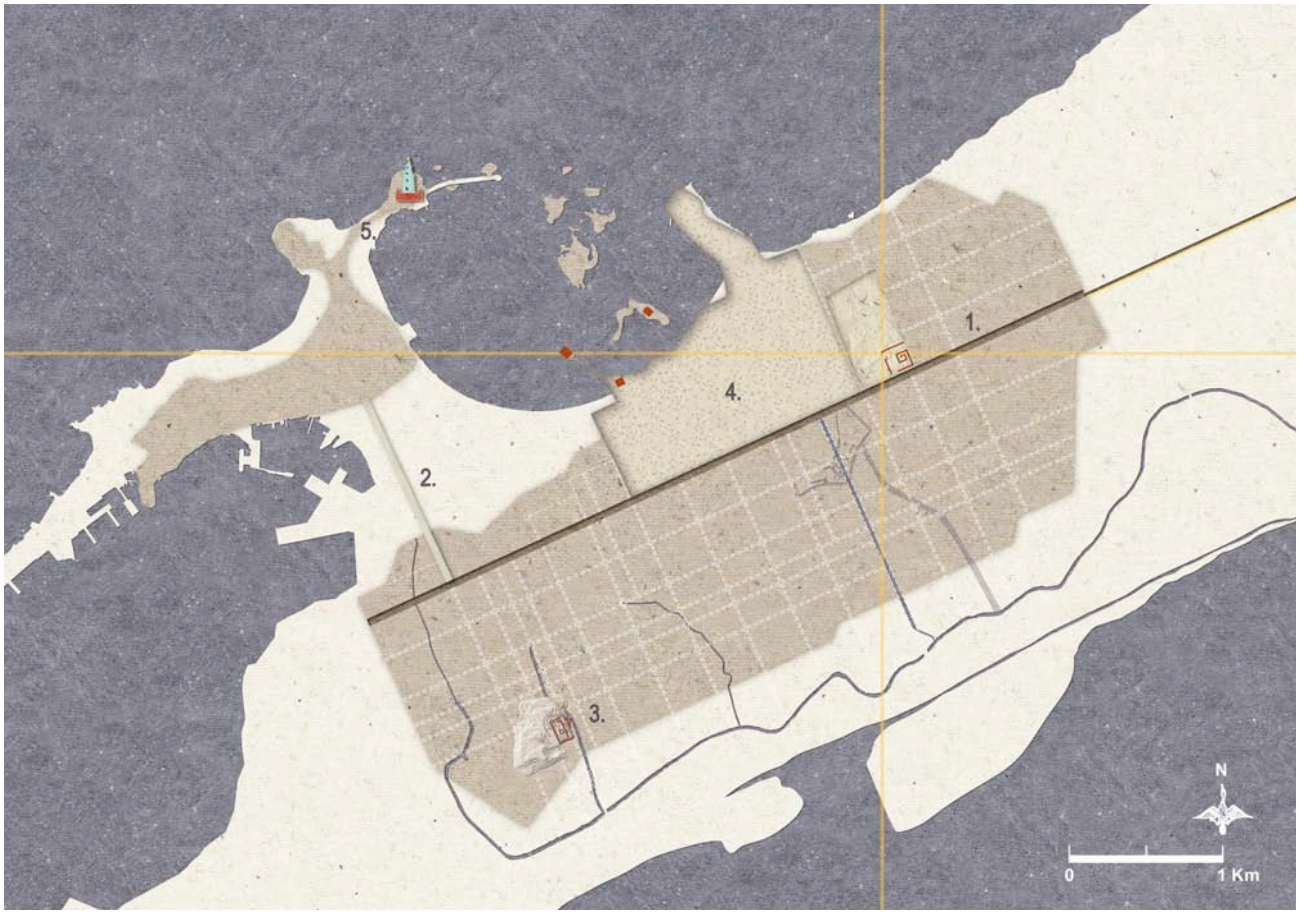

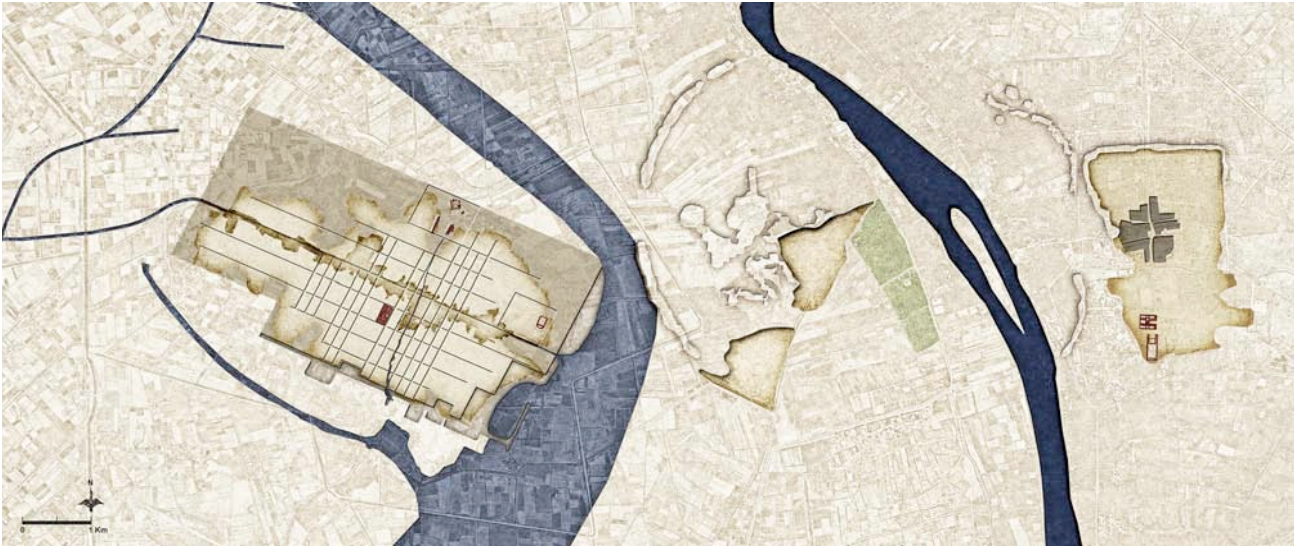

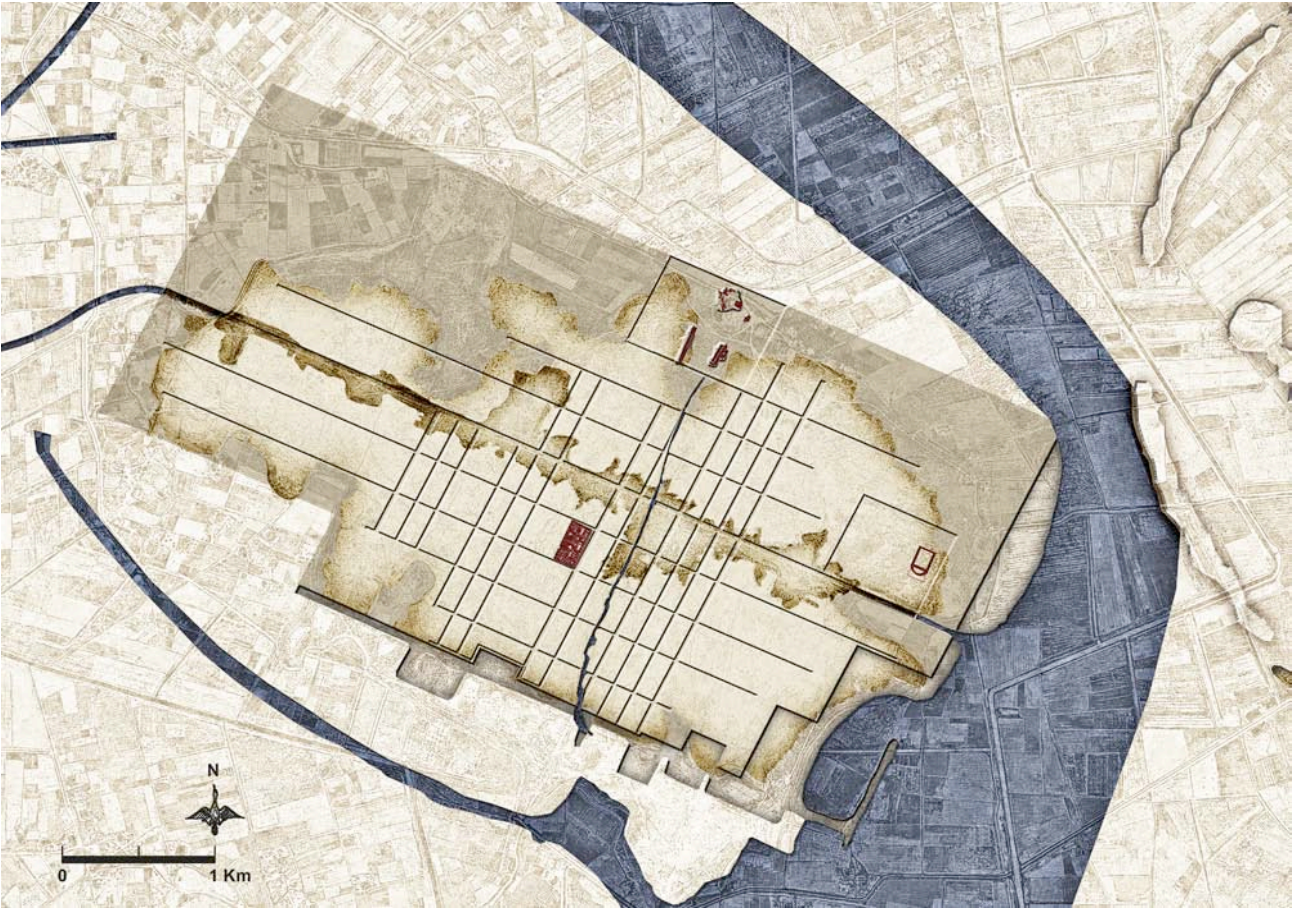

**Captions**

FIG 1 Alexandria, the original town plan superimposed on the reconstruction by Mahmoud-Bey (1866) (Courtesy of CEAlex Archives J-Y. Empereur).

FIG 2 The topography of Ancient Alexandria: traces, fragments and strata. Emphasis is given to the structures of the ancient *forma urbis* still identifiable in the city today. In additon the contour line that presumably marked the boundaries of the ancient city and parts of the coast now submerged (@ Luisa Ferro).

FIG 3 A photograph of the early 19$^{th}$ century showing the Canopic road, looking west (courtesy C. Pallini).

FIG 4 The position of the Caesareum in relation with the pre-existing grid and the sea (@ Luisa Ferro).

FIG 5 The original town plan superimposed on the plan of the Alexandria area from the city to Canopus/Abukir (from *Description de L'Egypte* (1803); courtesy of CEAlex Archives J-Y. Empereur).

FIG. 6 Alexandria, reconstruction scheme of the original town plan. 1) Canopic road 2) Heptastadion 3) Serapeum 4) Imperial palace 5) Isle of Pharos (@ Luisa Ferro).

FIG. 7 The topography of Seleucia on the Tigris and Ctesiphon superimposed on the contemporary aerial view. Emphasis is given to the structures still identifiable in the territory today, and to the old course of Tigri river. References: *Map of the Ruins of Ctesiphon and Seleucia* 1932 in The Metropolitan Museum Art Bulletin 8, vol. 27 (@ Luisa Ferro).

FIG. 8 Plan of Seleucia on the Tigris with excavation areas and reconstruction scheme of the original town plan. References: Michigan Univeristy and Toledo and Cleveland Museum excavations (1927-1936); Missione italiana in Iraq del Centro Ricerche Archeologiche e Scavi di Torino per il Medio Oriente e l'Asia (1964-1989) (@ Luisa Ferro).

---

1 Because of the superposition of the modern town, it is not easy to get the orientation of the Alexandria grid with a precision better than ½°, which, however, is definitively sufficient for the analysis presented here. To obtain such an accuracy we have compared several existing measures with measures obtained with a precision magnetic compass and corrected for magnetic declination.

2 This calculation has been done by several authors; we have independently controlled it as well. All astronomical data in the present paper have been elaborated using the software @StarryNight Pro 6.0.

3 In 331 BC the first new moon after the summer solstice occurred on July 9, thus giving July 14 for *Hecatombaeon* 6. This day corresponds to the sun rising about half-way between the summer solstice and the orientation of the grid.